\documentclass[
    ,final            % use final for the camera ready runs
  ]
  {aipproc}

\layoutstyle{6x9}

\begin{document}

\title{Electromagnetic transitions for $A$=3 nuclear systems}

\author{L.E. Marcucci}{
  address={Department of Physics, University of Pisa, I-56100 Pisa, Italy} 
,altaddress={INFN, Sezione di Pisa, I-56100 Pisa, Italy}
}
\author{M. Viviani}{
  address={INFN, Sezione di Pisa, I-56100 Pisa, Italy} 
,altaddress={Department of Physics, University of Pisa, I-56100 Pisa, Italy}
}
\author{R. Schiavilla}{
  address={Department of Physics, Old Dominion University, Norfolk, 
Virginia 23529, USA} 
,altaddress={Jefferson Lab, Newport News, Virginia 23606, USA}
}
\author{A. Kievsky}{
  address={INFN, Sezione di Pisa, I-56100 Pisa, Italy} 
,altaddress={Department of Physics, University of Pisa, I-56100 Pisa, Italy}
}
\author{S. Rosati}{
  address={Department of Physics, University of Pisa, I-56100 Pisa, Italy} 
,altaddress={INFN, Sezione di Pisa, I-56100 Pisa, Italy}
}

\begin{abstract}
 Recent advances in the study of $pd$ radiative 
capture in a wide range of center-of-mass energy 
below and above deuteron breakup threshold 
are presented and discussed. 
\end{abstract}

\maketitle

%%%%%%%%%%%%%%%%%%%%%%%%%%%%%%%%%%%%%%%%%%%%
%% MAINMATTER
%%%%%%%%%%%%%%%%%%%%%%%%%%%%%%%%%%%%%%%%%%%%

\section{Introduction}
\label{intro}

The electromagnetic transitions for $A$=3 nuclear systems 
have been extensively studied by several research groups
(see Ref.~\cite{Car98} for a review). The advantage of 
investigating these processes is that nowadays 
several methods can provide accurate bound- and 
scattering-state wave functions using realistic 
Hamiltonian models. Therefore, different models for the 
nuclear electromagnetic current operator can be tested 
with the large body of available experimental data.
In the present study, we concentrate our attention on the $pd$ radiative 
capture reaction in a wide range of center-of-mass energy ($E_{c.m.}$). 
This reaction was already studied below the deuteron breakup threshold 
by our group in Ref.~\cite{Viv00}, using  
pair-correlated hyperspherical harmonics (PHH)
wave functions  obtained from a realistic Hamiltonian model 
consisting of the Argonne $v_{18}$ two-nucleon~\cite{Wir95} and 
Urbana IX three-nucleon~\cite{Pud95} 
interactions (AV18/UIX). The nuclear electromagnetic current operator 
included, in addition to the one-body 
term, also two-body contributions, constructed 
following the method of Ref.~\cite{Ris85}, with the goal of 
satisfying the current conservation relation (CCR) with the AV18. 
However, within this method, the two-body terms 
originated from the momentum-dependent part of the AV18 were not strictly 
conserved. The $pd$ radiative capture observables were all quite well 
reproduced with the ``full'' model for the nuclear current operator, 
with the only exception of the deuteron tensor polarization 
observables $T_{20}$ and $T_{21}$. 
In the analysis of Ref.~\cite{Viv00}, it was concluded 
that these discrepancies might be due to the fact that the 
electromagnetic current operator satisfies the CCR 
with the nuclear Hamiltonian only approximately. 

In this work we present a new model for the nuclear 
current operator constructed so as to 
satisfy {\it exactly} the CCR with 
the AV18/UIX Hamiltonian model, and we test it in the study of 
the $pd$ radiative capture observables in a wide range of 
$E_{c.m.}$.
The model for the nuclear electromagnetic current operator is 
summarized in the following section. A detailed review 
will be given elsewhere~\cite{Marip}.

\section{The nuclear electromagnetic current operator}
\label{sec:em}

The nuclear electromagnetic current operator
can be written as sum of one- and 
many-body terms that operate on the nucleon degrees of freedom. 
The one-body operator is derived from the non-relativistic reduction
of the covariant single-nucleon current, by expanding
in powers of $1/m$, $m$ being the nucleon mass~\cite{Car98}.
To construct the two-body current operator,
it is useful to adopt the classification scheme of Ref.~\cite{Ris89},
and separate the current into model-independent (MI)
and model-dependent (MD) parts. 
The MD two-body current
is purely transverse and therefore is
un-constrained by the CCR. It is taken
to consist of the isoscalar $\rho\pi\gamma$
and isovector $\omega\pi\gamma$ transition currents, as well as the
isovector current associated with excitation of intermediate
$\Delta$ resonances as in Ref.~\cite{Viv00}.
The MI two-body currents have longitudinal
components and have to satisfy the CCR with the two-nucleon 
interaction. 
The MI terms arising from the momentum-independent terms 
of the AV18 two-nucleon interaction have been constructed 
following the standard procedure of Ref.~\cite{Ris85},
hereafter quoted as meson-exchange (ME) scheme.
It can be 
shown that these two-body current operators satisfy {\it exactly} the CCR
with the first six operators of the AV18. 
The two-body currents arising from the spin-orbit components of 
the AV18 could be constructed using again ME
mechanisms~\cite{Car90}, 
but the resulting currents turn out to be not strictly conserved. 
The same can be said of those currents 
from the quadratic momentum-dependent components of the AV18, 
if obtained, as in Ref.~\cite{Viv00}, 
by gauging only the momentum operators, 
but ignoring the implicit momentum dependence which comes through
the isospin exchange operator (see below).
Since our goal is to construct MI two-body currents 
which satisfy {\it exactly} the CCR with the complete 
AV18 two-nucleon interaction, 
the currents arising from the momentum-dependent 
terms of the AV18 interaction have been obtained following the 
procedure of Ref.~\cite{Sac48}, which will be quoted 
as minimal-substitution (MS) scheme.
The main idea of this procedure, 
fully reviewed in Ref.~\cite{Marip}, is that 
the isospin operator ${\bf \tau}_i\cdot{\bf \tau}_j$ 
is formally 
equivalent to an implicit momentum dependence~\cite{Sac48}. In fact, 
${\bf \tau}_i\cdot{\bf \tau}_j$ can be expressed in terms of the 
space-exchange operator ($P_{ij}$)  
using the formula $
  {\bf \tau}_i\cdot{\bf \tau}_j = -1-
  (1+{\bf \sigma}_i\cdot{\bf \sigma}_j) P_{ij}$, 
valid when operating on antisymmetric wave functions. 
Note that the operator $P_{ij}$ is defined as 
$P_{ij}=   {\rm e}^{{\bf r}_{ji}\cdot{\bf\nabla}_i
  + {\bf r}_{ij}\cdot{\bf\nabla}_j}$, where 
the ${\bf \nabla}$-operators do not act on the 
vectors ${\bf r}_{ij}={\bf r}_i-{\bf r}_j=-{\bf r}_{ji}$ 
in the exponential.
In the presence of an electromagnetic
field, minimal substitution is performed both in the 
momentum dependent terms of the two-nucleon interaction and in the  
space-exchange operator $P_{ij}$. 
The resulting current operators are then obtained with standard 
procedures~\cite{Marip,Sac48}. 
Explicit formulas can also be found in Ref.~\cite{Marip}. 

Both the ME and the MS schemes 
can be generalized to calculate the three-body current operators 
induced by the three-nucleon interaction (TNI).
Here, these three-body currents have been constructed within the ME scheme 
to satisfy the CCR with the Urbana-IX TNI~\cite{Pud95}. 
Details of the calculation can be found in Ref.~\cite{Marip}.

In summary, the present model for the many-body current operators 
retains 
the two-body terms obtained within the ME scheme from the 
momentum-independent 
part of the AV18, those ones obtained within the MS 
scheme from the momentum-dependent part of the AV18, the MD
terms quoted above, and the three-body terms obtained 
within the ME scheme from the UIX TNI. 
Thus, the full current operator satisfies 
{\it exactly} the CCR with the AV18/UIX nuclear Hamiltonian. 
In contrast, the model of Ref.~\cite{Viv00} 
retains only two-body currents, all of them obtained within the 
ME scheme.

\begin{figure}
\includegraphics[height=.3\textheight]{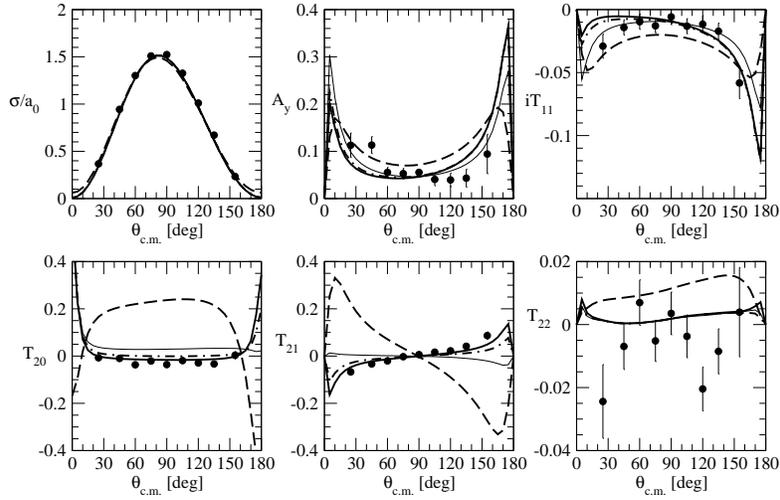}
\caption{Differential cross section, proton vector analyzing power, 
and the four deuteron tensor polarization observables 
for $pd$ radiative capture at $E_{c.m.}=$2.00 MeV, obtained with the 
AV18/UIX Hamiltonian model. 
See text for the explanations of the different curves.
The experimental data are from Ref.~\protect\cite{Smi99}.}
\label{fig:obs.2.00}
\end{figure}

\begin{figure}
\includegraphics[height=.3\textheight]{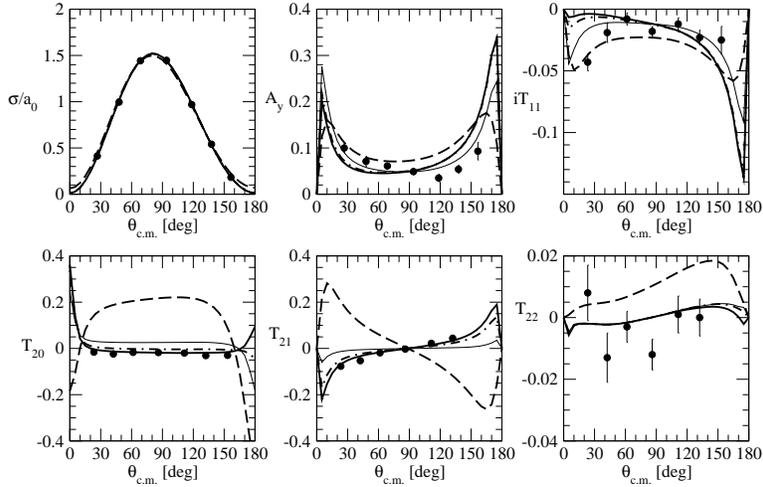}
\caption{Same as Fig.~\protect\ref{fig:obs.2.00}, 
but at $E_{c.m.}=$3.33 MeV. 
The experimental data are from Ref.~\protect\cite{Goe92}.}
\label{fig:obs.3.33}
\end{figure}

\section{Results}
\label{sec:res}

\begin{figure}
\includegraphics[height=.3\textheight]{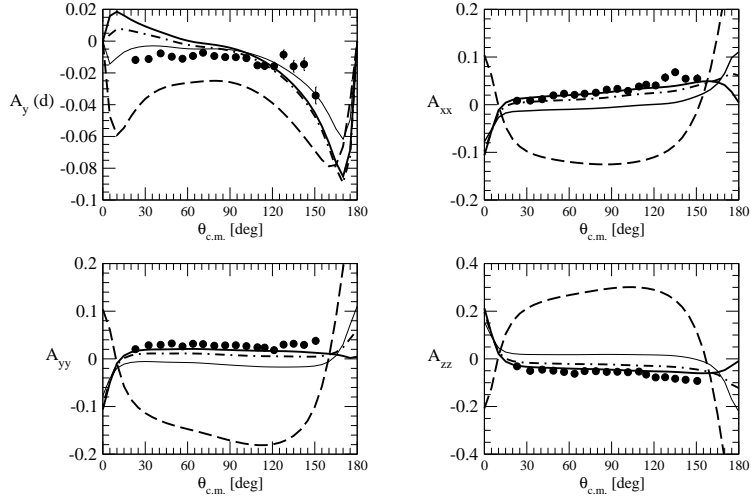}
\caption{Deuteron vector and tensor polarization observables 
at $E_{c.m.}=$5.83 MeV. 
The experimental data are from Ref.~\protect\cite{Aki01}.}
\label{fig:obs.5.83}
\end{figure}

\begin{figure}
\includegraphics[height=.3\textheight]{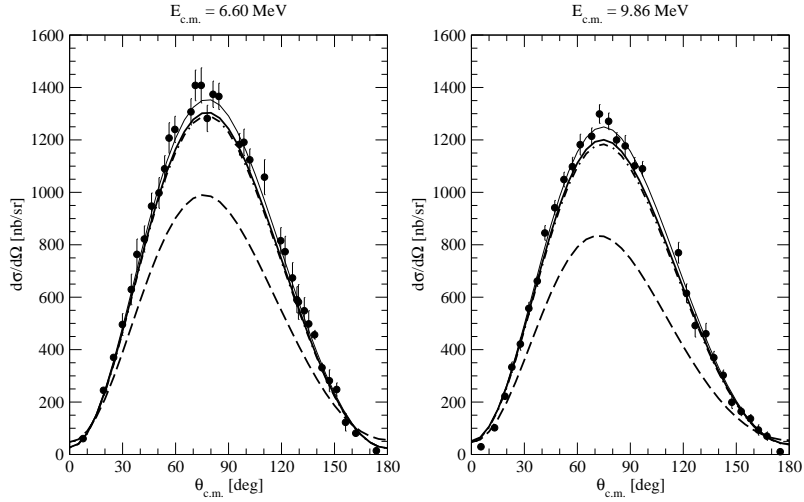}
\caption{Differential cross section for $pd$ radiative capture 
at $E_{c.m.}=$6.60 MeV (left panel) and 9.86 MeV (right panel). 
The experimental data are from Ref.~\protect\cite{Bel70}.}
\label{fig:obs.6.60-9.86}
\end{figure}

The theoretical predictions of the $pd$ radiative capture observables at 
$E_{c.m.}=$2.00--9.86 MeV obtained with the AV18/UIX 
Hamiltonian model are compared with the available experimental 
data in Figs.~\ref{fig:obs.2.00}--~\ref{fig:obs.6.60-9.86}. 
In all the figures, the dashed, dotted-dashed and thick-solid 
curves correspond
to the calculation with one-body only, with one- and 
two-body, and with one-, two- and three-body 
currents. Also shown (thin-solid lines) are the results obtained 
with the model for the nuclear current operator 
of Ref.~\cite{Viv00}. From inspection of the figures, 
we can conclude that: (i) the present ``full'' model for the nuclear 
electromagnetic current operator provides 
an overall nice description for 
all the observables, with the only exception of the 
deuteron vector analyzing powers i$T_{11}$ 
and $A_{y}(d)$ at small center-of-mass angles. The origin 
of these discrepancies is currently under investigation. 
A possible explanation could be a deficiency in the used model for the  
TNI, in particular in the absence of three-nucleon spin-orbit 
terms~\cite{Kie99}.
(ii) Some 
small three-body current effects are noticeable, especially in 
the deuteron tensor polarization observables $T_{20}$, $T_{21}$ 
at $E_{c.m.}=$2.00 and 3.33 MeV and 
$A_{xx}$, $A_{yy}$ and $A_{zz}$ at $E_{c.m.}=$5.83 MeV. 
This is an indication of the fact that 
if a Hamiltonian 
model with two- and three-nucleon interactions is used, then the 
model for the nuclear current operator should include the corresponding 
two- and three-body contributions.
(iii) The results obtained with the model of Ref.~\cite{Viv00} 
are in strong disagreement with the data for the 
deuteron tensor polarization observables. This is a consequence of the 
fact that in Ref.~\cite{Viv00} the CCR is only approximately satisfied. 
However, this ``old'' model seems to provide a better description for the 
deuteron vector polarization observables i$T_{11}$ and $A_y(d)$. 

\section{Summary and Outlook}
\label{sec:sum}

We have reported new calculations for $pd$ radiative capture 
observables in a wide range of $E_{c.m.}$ below and above deuteron 
breakup threshold. 
These calculations use accurate bound and 
scattering state wave functions obtained with the PHH technique
from the Argonne $v_{18}$ two-nucleon and Urbana IX three-nucleon 
interactions. The model for the electromagnetic 
current operator includes one-, two- and three-body 
components, constructed so as to 
satisfy {\it exactly} the CCR with the given 
Hamiltonian model. An overall nice description of all the 
observables has been achieved, with the only exception 
of the deuteron vector analyzing powers at small angles. 
A systematic comparison between theory and experiment for 
the $pd$ radiative capture in a wider range of $E_{c.m.}$ 
is currently underway~\cite{Marip}.

The work of R.S. was supported by the U.S. DOE Contract 
No. DE-AC05-84ER40150, under which the Southeastern Universities 
Research Association (SURA) operates the Thomas Jefferson National 
Accelarator Facility.


\begin{thebibliography}{}
%
\bibitem{Car98} Carlson, J., and Schiavilla, R., 
                {\it Rev.\ Mod.\ Phys.}, \textbf{70}, 743 (1998).
%
\bibitem{Viv00} Viviani, M., Kievsky, A., Marcucci, L.E., Rosati, S., 
		and Schiavilla, R., 
                {\it Phys.\ Rev.\ C}, \textbf{61}, 064001 (2000).
%
\bibitem{Wir95} Wiringa, R.B., Stoks, V.G.J., and Schiavilla, R., 
                {\it Phys.\ Rev.\ C}, \textbf {51}, 38 (1995).
%
\bibitem{Pud95} Pudliner, B.S., Pandharipande, V.R., Carlson, J., 
		and Wiringa, R.B.,
                {\it Phys.\ Rev.\ Lett.}, \textbf {74}, 4396 (1995).
%
\bibitem{Ris85} Riska, D.O.,
                {\it Phys.\ Scr.}, \textbf {31}, 107 (1985);
		{\it Phys.\ Scr.}, \textbf {31}, 471 (1985).
%
\bibitem{Marip} Marcucci, L.E., Viviani, M., Schiavilla, R., 
                Kievsky, A., and Rosati, S., in preparation.
%
\bibitem{Ris89} Riska, D.O., 
                {\it Phys.\ Rep.}, \textbf{181}, 207 (1989).
%
\bibitem{Car90} Carlson, J., Riska, D.O., Schiavilla, R., and Wiringa, R.B., 
                {\it Phys.\ Rev.\ C}, \textbf{42}, 830 (1990).
%
\bibitem{Sac48} Sachs, R.G., {\it Phys.\ Rev.}, \textbf{74}, 433 (1948);
                Nyman, E.M., {\it Nucl.\ Phys.}, \textbf{B1}, 535 (1967).
%
\bibitem{Kie99} Kievsky, A., 
                {\it Phys.\ Rev.\ C}, \textbf{60}, 034001 (1999).
%
\bibitem{Smi99} Smith, M.K., and Knutson, L.D.,
                {\it Phys.\ Rev.\ Lett.}, \textbf{82}, 4591 (1999).
%
\bibitem{Goe92} Goeckner, F., Pitts, W.K., and Knutson, L.D.,
                {\it Phys.\ Rev.\ C}, \textbf{45}, R2536 (1992).
%
\bibitem{Aki01} Akiyoshi, H., {\it et al.}, 
                {\it Phys.\ Rev.\ C}, \textbf{64}, 043001 (2001).
%
\bibitem{Bel70} Belt, B.D., {\it et al.}, 
                {\it Phys.\ Rev.\ Lett.}, \textbf{24}, 1120 (1970).
%
\end{thebibliography}
\end{document}